\documentclass[11pt]{article}
\usepackage{graphicx, color}
\usepackage{tikz}
\usetikzlibrary{arrows}
\usepackage[sectionbib]{natbib}
\usepackage[colorlinks=true, urlcolor={violet}, citecolor={blue}, linkcolor={red}]{hyperref}
\usepackage{rotating}
\usepackage{verbatim,listings}
\usepackage{fullpage}
\usepackage{amssymb}
\usepackage[reqno]{amsmath}
\usepackage{setspace}
\usepackage{anysize}
\usepackage{lscape}
\usepackage{bm}
\usepackage{booktabs}
\usepackage{changepage}
\usepackage{tikz}
\def\citeapos#1{\citeauthor{#1}'s (\citeyear{#1})}
\definecolor{MyGreen}{cmyk}{1.0,0.0,1.0,0.2}

\newcommand{\bt}{\pmb{\theta}}

\newcommand{\Y}{\bm{\mathcal{Y}}}

\newcommand{\h}{\mathbf{h}}

\def\citeapos#1{\citeauthor{#1}'s (\citeyear{#1})}

\newlength\Colsep
\begin{document}

\vspace{2cm}

\title{Statistical Inference in Political Networks Research\footnote{Many thanks to Ramiro Berardo, Jan Box-Steffensmeier, Dino Christenson, Michael Heaney, and Philip Leifeld for helpful comments. Work on this chapter was supported by National Science Foundation awards SES-1357622, SES-1461493, SES-1514750, SES-1357606, and CISE-1320219 and by the Alexander von Humboldt Foundation.}}
\author{Bruce A. Desmarais\thanks{\footnotesize{
Bruce A. Desmarais is an Associate Professor of Political Science and an affiliate of the Institute for CyberScience at the Pennsylvania State University.}} and Skyler J. Cranmer\thanks{\footnotesize{Skyler J. Cranmer is the Carter Phillips and Sue Henry Associate Professor of Political Science at the Ohio State University. He is also an Alexander von Humboldt Fellow visiting the University of Konstanz.}}}
\date{}
\maketitle

\begin{abstract}
\noindent Researchers interested in statistically modeling network data have a well-established and quickly growing set of approaches from which to choose. Several of these methods have been regularly applied in research on political networks, while others have yet to permeate the field. Here, we review the most prominent methods of inferential network analysis -- both for cross-sectionally and longitudinally observed networks including (temporal) exponential random graph models, latent space models, the quadratic assignment procedure, and stochastic actor oriented models. For each method, we summarize its analytic form, identify prominent published applications in political science and discuss computational considerations. We conclude with a set of guidelines for selecting a method for a given application.
\end{abstract}
~\\
Appearing in {\em The Oxford Handbook of Political Networks.} Jennifer Nicoll Victor, Alexander H. Montgomery, and Mark Lubell, editors. Oxford University Press, 2017.

\doublespacing

\section{Introduction}

Networks are systems that exhibit complex interwoven structures. These systems can be illuminating to describe and visualize, but most research agendas eventually arrive at either precise hypotheses regarding endogenous network structures (e.g. the structure of connections within the outcome network of interest) or similarly precise hypotheses regarding exogenous factors but requiring adjustment for endogenous dependencies. The conventional toolkit for statistical inference is poorly suited to evaluating such hypotheses. The core problem is that the classical machinery of hypothesis testing was designed around the assumption that the dataset to be used is populated with independent replicates from a data generating process. Network data could not be further from this assumption. The observations, which are often organized as dyads, are individual components of a single broader system. Even if a researcher has several replicates of the system, they are often dependent (e.g., longitudinal network data). Because the interdependencies inherent in networks violate a core assumption upon which standard regression models are built, interest in networks has sparked a wave of methodological innovations aimed at statistical inference, modeling, and hypothesis testing with networks. Here, we review the most established of these methods, briefly discuss their structure and capabilities, and highlight selected applications of these tools in political science research. Obviously, a review chapter such as this cannot discuss the intricacies of each model exhaustively, but we hope to provide a general orientation to the set of techniques available and point the reader to where further details and examples may be had. 

\section{Cross-Sectional Models}

The basic approaches to statistical inference with networks are differentiated by the objectives of the researcher vis-a-vis the network structure and dependencies. The approaches that we review below permit researchers to test precise theories of network formation that involve both the effects of covariates on tie formation and relationships among ties themselves. We also present methods that permit inference when the researcher is uncertain regarding the forms of dependence that underlie tie formation in the network. Here, we provide a very brief overview of the three major methods of cross sectional network analysis and provide some discussion about their use in the political science literature. For a more detailed discussion of how these methods relate to each other, and for an illustrative application of all three to the same dataset, see \cite{Cranmer:2016}.

\subsection{The Exponential Random Graph Model (ERGM)}

We consider the Exponential Random Graph model (ERGM) \citep{Holland:1981,Wasserman:1996}, to be the canonical model for network data. The ERGM provides a modeling framework that can, just like conventional regression, accommodate the effects of covariates on the status of relationships, but unlike regression, can model the prominence and significance of structural dependencies such as reciprocity and transitivity. Furthermore, the ERGM has been extended to accommodate longitudinal relational data (i.e., time series of networks) \citep{Robins:2001,Hanneke:2010,Desmarais:2012physa} and networks with edge weights \citep{Wyatt:2010,Desmarais:2011plosone,Krivitsky:2011}. 

The mathematical form of the ERGM makes clear its potential as a general framework for statistical inference with networks.  The ERGM is a probability model defined on the support of possible adjacency matrices.  Let there be $n$ actors/nodes/vertices, among which which we represent dichotomous ties as $Y$, an $n\times  n$ matrix with $Y_{ij}=1$ if there is a tie from $i$ to $j$ and 0 otherwise. The probability of observing $Y$ according to an ERGM is
\begin{equation}
P(Y) = \frac{   \exp(   \sum_{j=1}^k \theta_jh_{j}(Y))   } {\sum_{Y^*\in \Y} \exp(   \sum_{j=1}^k \theta_jh_{j}(Y^*))}.
\label{pg}
\end{equation}
The flexibility of the ERGM for capturing covariate effects and interdependencies in networks arises from the general nature of the $h_j$. These are statistics -- functions applied to the adjacency matrix and optionally covariates -- that are theorized to influence the likelihood of observing a configuration of the adjacency matrix (e.g., a network with 30 triangles, 100 edges and 25 reciprocal dyads). The $\theta$ are parameters to be estimated that, similar to regression coefficients, give the effects of the network statistics on the likelihood of observing a particular instance of the adjacency matrix. The larger $\theta_j$, the higher the likelihood of observing subgraph configurations that contribute positively to $h_j$. The flexibility and grounding in theory come at a cost. An identifying assumption, one also common to the regression framework, is that the model is correctly specified. That is, it is assumed that every feature of a network configuration that is relevant to determining whether we will observe that network configuration is represented by an element of $h$. Misspecification can result in biased estimates of parameters as well as sub par performance of hypothesis tests.  A detailed review of the ERGM is provided by a number of sources including \cite{Goodreau:2008}, \cite{hunter2008ergm}, \cite{cranmer2011inferential},  and \cite{Lusher:2013}.

For applications of ERGM in political networks research see \cite{thurner2009}, \cite{lazer2010}, \cite{cranmer2011inferential}, \cite{gerber2013}, \cite{kirkland2014}, \cite{boxsteffensmeier2014}, \cite{song2015}.

Let us consider the example of \cite{boxsteffensmeier2014} in detail. Recently, scholars have begun to construct networks of interest groups based on cosigning the same amicus curiae brief.  Amicus curiae or ``friend of the court'' legal briefs are written by actors who are not party to a case but believe they can offer information relevant to it. One contribution of this work is that the network measure constructed from amicus curiae briefs is both coordinated and purposive. The process requires negotiation between cosignatories and agreement on the direction, argument and details of the brief. They find evidence of an increasingly transitive network resembling a host of tightly grouped factions and leadership hub organizations employing mixed coalition strategies. Egocentric networks of organizations show that three major theoretically posited coalition strategies are present in the data: ``lone wolves,'' who work alone; ``teammates,'' who work with cliques of varying sizes; and ``leaders,'' who pull together otherwise disparate groups. They further utilized business directories to gather attribute data on signers from the first decade of the millennium. In doing so, they find a broad range of industries, with the greatest number of signers in the Services Division of the Standard Industrial Classification (SIC). They also use the attribute data to test homophily hypotheses for various business characteristics in the ERGM. The paper finds that driving the network formation is assortative mixing based on industry area, budget, sales and membership characteristics. 

Substantively, the paper contributes to the discipline by shedding light on how to measure the networks of the vast expanse of special interests, how the networks have changed over time, and which characteristics are sought out as complements and which are considered threatening to cooperation. In addition it provides insights into more micro-level considerations about which groups are the most attractive partners and in what capacity. 

\subsection{The Latent Space Model (LSM)}

The latent space model, introduced by \citet{Hoff2002a} is latent variable model for tie formation in networks that exploits the near ubiquity of homophily in networks -- the tendency for nodes to tie to others that are similar on one or more attribute. Nodes are positioned in a $k$-dimensional latent space and the probability of a tie between any two nodes is inversely related to the distance between their positions in this latent space. Each dimension in the latent space can be thought of as an unmeasured attribute of nodes. For example, such attributes might be income or age if nodes are people; or level of authoritarianism and unemployment rate if nodes are countries.  Under the latent space model, the log odds of a tie between nodes $i$ and $j$ is:
\begin{align}
\eta_{i,j} &= \text{log odds}(y_{i,j} = 1| z_i, z_j, x_{i,j}, \alpha, \beta)\\
&= \alpha +\beta'x_{i,j} - | z_i - z_j|
\end{align}
\noindent  where $\alpha$ is an intercept term controlling the overall density of the network, $\beta$ is a vector of regression coefficients,  $x_{i,j}$ is vector of measured covariates, and $| z_i - z_j|$ is the euclidean distance between nodes in the latent space. Interpretation of the LSM is similar to interpreting logistic regression. The $\beta$ is interpreted as the change in the log odds of a tie given a one unit increase in the value of a covariate.

For applications of LSM in political networks research see \cite{ward2007}, \cite{kirkland2012}, \cite{krafft2012}, \cite{ward2013}, \cite{kirkland2014}, \cite{cao2014}.

To consider an example of such work in detail, let us examine \cite{kirkland2012}. Kirkland was primarily interested in the effect of an exogenous variable, the sharing of a constituency, on what on the network of cosponsorships in the US Congress. In terms of design, he used a natural experiment: during the 2000-2002 redistricting cycle, NC eliminated its small set of multi-member districts as a result of a court intervention, allowing him to compare the behavior of legislators in multi-member districts to those in single-member districts (since not all NC districts were multi-member), and to compare legislators from multi-member districts to themselves after the switch. His theory, being about exogenous predictors, did not provide specific hypotheses about the structure of the endogenous network dependencies, the making the proper specification of an ERGM difficult.  Treating the network dependencies as a nuisance, rather than an object of interest, he estimated a latent space model on each year in the sample. This setup allowed Kirkland to test the effects of the exogenous variable of interest on the network that while controlling for the interdependence known to be present. To select the dimensionality of the LSM, Kirkland increased the number of dimensions in the model until the new dimensions stopped producing significant gains in model fit; two dimensions as it should happen. 

Substantively, \cite{kirkland2012} demonstrates that sharing a constituency produces sizable increases in collaboration via cosponsorship. There is some research both in American politics and in the comparative literature that suggests that multi-member districts produce representatives who are likely to compete with one another to build a distinct ``brand'' from their district partner, and that these districts produce representatives who lack strong geographic ties to their constituents allowing them to adhere to party lines more frequently. Alternatively, Kirkland hypothesized that the advantages of working together to gain on policy outweighed the potential gains from adhering to party lines and building distinct individual brands and found that multi-member districts provide legislators with natural coalition partners, who stand to gain on policy through collaboration.

\subsection{The Quadratic Assignment Procedure (QAP)}

The quadratic assignment procedure (QAP) \citep{Krackardt1987} is a method of hypothesis testing that builds upon \citeapos{Mantel1967} permutation procedure and computes the statistical significance of parameter estimates when the dependent variable is itself a relational matrix (e.g., distance, correlation, network adjacency matrices). The QAP is a form of non-parametric permutation testing in which row and column shuffles of the dependent and independent variables are used to simulate the null condition, which results in a simulated null distribution of regression coefficients. 

To illustrate the basic methodology of QAP, we offer an example of bivariate permutation in Figure 1. Randomly shuffling rows and columns of the adjacency matrix (Y) maintains the structure of the network right down to the edges, but by shuffling the nodes, the relationship between Y and a hypothetical dyadic covariate X, is broken. In the QAP, some measure of relationship (e.g., correlation coefficient, regression) between Y and X is calculated on the un-permuted data, then the same measure is calculated on many permutation based replications of the data. A two-tailed $p$-value to assess the null hypothesis that there is no relationship between Y and X can be calculated as the proportion of relationships calculated under permutations that are at least as large, in magnitude, as the value calculated on the observed data.

\begin{figure}
\centering
\noindent\begin{minipage}{.5\textwidth}
\begin{minipage}[c][5cm][c]{\dimexpr0.5\textwidth-0.5\Colsep\relax}
 \begin{center}
    {\bf {\em Y}} \\ \vspace{.2cm}
\begin{tabular}{|c|c|c|c|}
\hline  & A & B & C \\ 
\hline  A& 0 & 1 & 1  \\ 
\hline  B& 1 &  0 & 0 \\ 
\hline  C & 0 & 1 & 0 \\ 
\hline 
\end{tabular} \\
$\vdots$ \\
\begin{tabular}{|c|c|c|c|}
\hline  & B & C & A \\ 
\hline  B & 0 & 0 & 1 \\ 
\hline  C & 1 & 0 &0  \\ 
\hline  A &1  & 1 & 0 \\ 
\hline 
\end{tabular} 
    \end{center} \end{minipage}\hfill
\begin{minipage}[c][5cm][c]{\dimexpr0.5\textwidth-0.5\Colsep\relax}
\begin{center}
{\bf {\em X}} \\ \vspace{.2cm}
\begin{tabular}{|c|c|c|c|}
\hline  & A & B & C \\ 
\hline  A& 0 & 3 & -2  \\ 
\hline  B& 1.2 &  0 & -4 \\ 
\hline  C & 0 & 6 & 0 \\ 
\hline 
\end{tabular} \\
$\vdots$ \\
\begin{tabular}{|c|c|c|c|}
\hline  & A & B & C \\ 
\hline  A& 0 & 3 & -2  \\ 
\hline  B& 1.2 &  0 & -4 \\ 
\hline  C & 0 & 6 & 0 \\ 
\hline 
\end{tabular} \\
\end{center}
\end{minipage}%
\end{minipage}
\caption{Illustration of bivariate QAP. Example of dependent network (i.e., Y) permutation.}
\end{figure}

Using QAP with three or more values is more complicated than using either Y or X permutation, which are equivalent in the bivariate case. Suppose there are three relational variables -- the dependent network Y, the independent variable of interest X, and a potential confounder Z, for which we would like to control in a regression calculating the effect of X on Y. Figure 2 depicts the causal diagram of Z inducing a spurious correlation between X and Y. Consider Y permutation, in which the rows of Y are permuted to reflect the null hypothesis of no relationship between Y and X. Doing this also breaks the relationship between Z and Y, which amounts to assuming away one leg in the confounder diagram. Now consider X permutation, in which just X is shuffled. This would break the relationship between Z and X, also assuming away a leg in the confounding diagram.

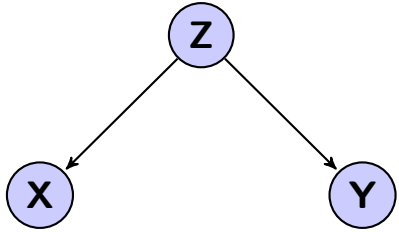
\begin{figure}

\tikzstyle{vertex}=[circle,fill=black!25,minimum size=15pt,inner sep=0pt]
\tikzstyle{selected vertex} = [vertex, fill=red!24]
\tikzstyle{edge} = [draw,thick,-]
\tikzstyle{weight} = [font=\small]
\tikzstyle{selected edge} = [draw,line width=5pt,-,red!50]
\tikzstyle{ignored edge} = [draw,line width=5pt,-,black!20]

\begin{center}
 \begin{tikzpicture}[->,>=stealth',shorten >=1pt,auto,node distance=3cm,
  thick,main node/.style={circle,fill=blue!20,draw,font=\sffamily\Large\bfseries}]

  \node[main node] (Z) {Z};
  \node[main node] (Y) [below right of=Z] {Y};
  \node[main node] (X) [below left of=Z] {X};
a
  \path[every node/.style={font=\sffamily\small}]
    (Z) edge node {} (X)
     (Z) edge node {} (Y);
\end{tikzpicture}
\end{center}

\caption{Causal diagram illustrating the role of a confounder (Z) in creating a spurious correlation between X and Y.}
\end{figure}

\cite{Dekker2007} propose an algorithm, termed double semi-partialing, that is intended to balance between X and Y permutation, and they show performs better than X or Y permutation. Let $\hat{\delta}Z$ be the estimate from $Y= \delta Z+E$. Define $\hat{\epsilon}_{XZ} = X - \hat{\delta}Z$.  Define $\pi(\hat{\epsilon}_{XZ})$ to be a matrix permutation of $\hat{\epsilon}_{XZ}$.  The non-parametric null distribution of $\beta$ is derived by estimating $Y = \pi(\hat{\epsilon}_{XZ})\beta+\delta Z+E$. QAP is a very useful methodology when the researcher is uncertain regarding what dependencies to include in ERGM or the latent space model does not provide a good fit.

For applications of QAP in political networks research see \cite{mizruchi1990},
\cite{Weible2005}, \cite{bochsler2009}, \cite{shrestha2009}, \cite{henry2011}, \cite{lee2012}, \cite{heaney2012},  \cite{desmarais2015b}, \cite{Andrew2015}, \cite{Chen2015}. 

Considering the work of \cite{henry2011} as an example, the authors were interested in understanding the conditions under which party activists are members of the same associations and particularly in assessing the extent to which co-partisanship affected this likelihood.  Drawing upon surveys they conducted at the 2008 Democratic and Republican National Conventions, \cite{henry2011} estimated models predicting organizational co-membership as a function of co-partisanship.  They also controlled for ideology, sex/gender, race/ethnicity, age, educational attainment, income, and religious participation.  The empirical analysis is particularly interesting given our discussion above, but cause the authors specified three types of models: logistic regression, the EQP, and an ERGM.  Results from all three models showed that being in the same party is a strong positive predictor of being in the same organization, even after controlling for alternative explanations for co-membership.  In fact, only 1.78\% of co-memberships crossed party lines: the association network is strongly polarized by party.  The results illustrate one of the many barriers to partisan polarization: party activists rarely come into contact with activists from the other party in their civic and political associations.

\section{Dynamic Models}

The statistical analysis of longitudinal networks is a smaller literature than that for cross-sectional networks. Two models that build on the ERGM, however, have been applied with some frequency in political science. Here, we discuss each model and their applications to political science briefly. For a direct contrast and comparison of the two models, see \cite{desmarais2012micro} and  \cite{leifeld2014comparing}

\subsection{The Temporal ERGM (TERGM)}
The temporal exponential random graph model (TERGM) is a straightforward longitudinal extension of the ERGM. The defining feature of the TERGM vis-a-vis the ERGM is that the statistics in the TERGM model the ways in which past realizations of the network influence the current network.

To modify the ERGM for modeling how the network $N$ at time $t$, denoted $N^t$, depends on previous networks $K \in \{1,\hdots, T-1\}$, we simply introduce the lagged networks into the $\h$. 
\begin{equation}\label{tergm}
P(Y^{t}| Y^{t-K}, \ldots, Y^{t-1}, \bt) = \frac{\exp(\bt^\top \h(Y^{t}, Y^{t-1}, \ldots, Y^{t-K}))}{c(\bt, Y^{t-K}, \ldots, Y^{t-1})}.
\end{equation}
The specification of $K$ is important because $Y^{t}$ networks in the time series preceding time point $K+1$ cannot be modeled with the correct model.

Equation~\ref{tergm} specifies a TERGM for a single time point, $Y^{t}$. The joint likelihood of the time series of networks spanning times $K+1$ and $T$ is derived as the product of the conditional probabilities -- conditioning on past networks:
\begin{equation}\label{tergm-pooled}
P(Y^{K+1},\ldots, Y^{T} | Y^{1},\ldots,Y^{K}, \bt) = \prod_{t=K+1}^{T} P(Y^{t}|Y^{t-K}, \ldots, Y^{t-1}, \bt).
\end{equation}
A TERGM can be applied to just a single network, which is equivalent to using ERGM with statistics involving one or more lagged networks, or a time series of networks with dozens or even hundreds of networks. For more detailed methodological treatment of the TERGM, see \cite{hanneke2010discrete}, \cite{cranmer2011inferential}, and \cite{desmarais2012statistical}. The bootstrap pseudo likelihood methods introduced in \cite{desmarais2012statistical} permit computationally efficient application to long time series.

Dependence on lagged networks can be incorporated in a fashion similar to including a covariate in a cross-sectional ERGM. For example, reciprocation is a nearly ubiquitous process in social networks. However, in time-stamped data, it is not clear whether the reciprocation of a tie will occur within one time period -- especially if ties take a long time to form (e.g., links in an annually measured scientific paper citation network) or if time periods are notably short (e.g., an e-mail network with hourly time stamps). To observe reciprocity, we may need to measure it with a lag. Single-period delayed reciprocity is modeled as:
\begin{equation}\label{lag-recip}
h_r(Y^{t}, Y^{t-1}) = \sum_{ji} Y^{t}_{ij}Y^{t-1}_{ji}.
\end{equation}
The lagged effects in a TERGM specification -- the components of $\h$ -- can vary in terms of their order, from 1 to $K$. For example, if $K=4$ we could model star-formation processes that aggregate over four periods, but also limit reciprocity effects to a single period. See \cite{morris2008specification} for an extensive review of endogenous effect specification.

For applications of the TERGM in political networks research, in which myriad examples of lagged endogenous effects can be found, see, e.g., \cite{Cranmer:2012}, \cite{Cranmer:2012ii}, \cite{Almquist:2013}, \cite{Clark2013}, \cite{Corbetta2013}, \cite{Cranmer:2014son}, \cite{Masket2014}, and \cite{ingoldforthcoming}.

Consider the work of \cite{ingoldforthcoming} as an example. Dozens of policy network case studies in continental Europe have used almost identical survey questionnaires to collect data over the last three decades. In addition to data on  collaboration, information exchange, and a variety of actor attributes, respondents were usually asked whom they would judge as particularly important in the rest of the network. While this is obviously a network relation, earlier studies aggregated it into a node attribute in order to measure an actor's overall importance in the policy network. \cite{ingoldforthcoming}exploited the fact that this is a network relation and modeled it using ERGMs for the cross-sectional datasets and TERGMs for the longitudinal data set. Specifically, they used four different datasets from different countries (institutional settings), levels (local to national), and policy stages (implementation and policy formulation decision making) to assure validity of results across contexts (and indeed they mostly are). In terms of specification, they estimated an ERGM of their second temporal observation conditional on the first, thus modeling the changes between periods one and two. 
The TERGM examined whether changes in the betweenness centrality of actors in the collaboration network could explain their perception as being important at the second time point. In other words, it is not simply collaboration centrality that makes one important; it is not simply collaboration ties with alters that make one regard that other organization as important; and it's not simply one's institutional role that makes one important (e.g., being a decision maker)---it's also that becoming more central over time makes alters perceive an ego as being an important player in the policy process. As the predictor of interest was betweenness centrality,  this reflects the ability to tie different communities in the network together.

\subsection{Stochastic Actor Oriented Models (SAOMs)}

The stochastic actor oriented model (SAOM) \citep{Snijders:2010} is a statistical model for longitudinal network data that has seen considerable application in political networks research and can be considered an alternative to TERGM. The SAOM is a structurally detailed model that is focused on interpretation at the node level. The SAOM and TERGM are similar in many underlying facets. Indeed, the equilibrium distribution of the SAOM is an exponential family random graph \citep{Snijders:2001}.  The ``actor orientation'' in the SAOM begins with an assumption about how the network changes. First, ties change one at a time. An actor is selected at random according to a Poisson process. The actor can then choose to leave its tie profile unchanged or change a single outgoing tie.  The rates that parameterize the Poisson processes can be actor-specific, modeling how tie volatility varies based on actor-level covariates. Let $Y^{(ij)}$ represent the network in which  element $ij$ of the adjacency matrix is toggled (changed from 0 to 1 or from 1 to 0). If selected to consider a tie change, the probability that actor $i$ changes its tie status with $j$ is proportional to $\exp(   \sum_{h=1}^k \theta_h \Gamma_{ih}(Y^{(ij)})) $, where  $\sum_{h=1}^k \theta_h \Gamma_{ih}(Y^{(ij)})$ is the actor ``objective function'' \citep{Snijders:2010}. Like the ERGM, the building blocks are regression coefficient style parameters $\theta$ and network statistics $\Gamma()$. The SAOM objective function is a dot product of network statistics that change with the ties sent by actor $i$ and real-valued parameters.

An added feature of the SAOM relative to TERGM is that it offers the option of specifying a dynamic model for  time-dependent actor attributes termed ``behavior.'' The behavior equation includes both other node covariates and functions of the network designed to evaluate how the network influences behavior. To develop the idea of the behavior model, first consider a regression for a node attribute at time $t$, which includes as covariates other node attributes as well as node-level statistics calculated on the network at time $t-1$ (e.g., node centrality, the average attribute value of a node's partners in the network). Instead of directly measuring the network at $t-1$ in order to model node attributes at time $t$, intermediate changes in both the network and behavior are simulated between times $t$ and $t-1$. The model for the change in behavior between times $t-1$ and $t$ includes an iterative process of change between the network and behavior -- changes that are simulated/inferred, as they are not observed \citep{Steglich:2010}.  This simulation of intermediate changes is the modeling feature that differentiates TERGM and SAOM. The canonical form of data for which the SAOM is designed is network snapshot data, in which the ties in a network are sampled at different points in time, and the researcher does not know how the network changed between snapshots. The changes are simulated according to the detailed structural assumptions in the SAOM. Finally, one should also note the widespread misconception that this coevolutionary model can identify separate influences for homophily and influence, two forces that are usually confounded in network research. As \cite{Snijders:2010} point out, the SAOM can only make this differentiation in extremely limited cases and the model is not generally capable of this differentiation in applied research contexts. 

Political science applications of SAOM include \cite{berardo2010}, \cite{fischer2012}, \cite{fischer2013}, \cite{berardo2013}, \cite{kinne2014}, \cite{Manger2014}.

As a last example from the literature, consider \cite{berardo2013}. \cite{berardo2013} studies perceptions of procedural fairness in self-organized networks of stakeholders in five U.S. estuaries where environmental problems are common. How members of a group perceive fairness in the decision-making processes is a key variable to study for researchers interested in the study of collaborative processes, given that those processes where procedural fairness is low usually are less likely to be sustained in time. Berardo finds that participation in networks does not alter perceptions of procedural fairness. Instead, it is perceptions of procedural fairness what impacts participation in networks, with homophily seeming to drive the process. In other words, similar perceptions of fairness is a strong predictor of the establishment of ties among actors. Berardo claims that this is an important result because it hints to the fact that, in the absence of high levels of procedural fairness, the network is likely to be fragmented into groups of ``winners'' and ``losers'' of the decision-making process, which may make long-term collaboration at a systemic level hard to sustain.

\section{Selecting an appropriate approach}

The methods described above differ along a couple of important dimensions. These dimensions include (1) the role of structural network theory in developing a model specification, and (2) the degree to which interdependence is accounted for through the model parameters vs through the hypothesis testing framework. Considering how a given application aligns along these dimensions, the researcher can determine which of these methods is most appropriate. 

The first major consideration in selecting a model for a network regards whether one of the objectives of the research is to conduct inference regarding precise patterns of interdependence among the ties in the network. For example, \cite{song2015} state several hypotheses regarding interdependence between ties in networks of political discussion. These include popularity (the tendency to send ties to nodes with high in-degree, activity (the tendency for nodes with high out degree to send new ties at a higher rate) and transitivity. When the researcher has precise hypotheses regarding dependence among the ties, the only models that we have discussed that can be used are those based on the ERGM -- ERGM, TERGM, and SAOM -- the latter two in the case of longitudinal networks. The LSM, though useful for the purpose of exploring the structure of the network that is not explained through the use of exogenous node and dyad covariates, cannot be used to test hypotheses regarding the interdependence among ties. QAP represents a method of hypothesis testing that is robust to the presence of interdependence among the ties in the network, but there is no way to use QAP results in diagnosing the forms of interdependence that exist among the ties.

If the researcher is only interested in testing the effects of covariates, but would like to account for network dependence, the second important consideration is the choice between parametric and non-parametric methods. Both the ERGM and LSM use precise parameterizations to model away the interdependence for which the covariates do not account. In the ERGM, the researcher must specify the exact network statistics that correspond to the forms of interdependence, a process which may sometimes be burdensome. In the LSM, the interdependence is absorbed by a large latent parameter space of unmeasured node attributes, along which nodes are assumed to be homophilous. 

With both the LSM and ERGM, it is possible that the researcher will fail to specify a model that results in the good fit required to draw inferences with a parametric model. As noted above, specification of an ERGM requires precise theory regarding the  forms of interdependence in a network. The LSM demands less in terms of precise assumptions regarding structural dependencies, but it does impose some limitations. The dependence structures in the LSM must conform closely to a metric space in which the probability of a tie is inversely proportional to distance in the space. This creates two limitations on the LSM in accounting for dependence. These limitations derive from the properties of metric spaces \citep{bryant1985}. The first property is {\em symmetry}, which means that the distance from $i$ to $j$ is equivalent to the distance from $j$ to $i$. If the probability of a tie is linked to these symmetric distances in the latent space, then the latent space model implies that ties exhibit a tendency towards reciprocity. Symmetric dependence would be inappropriate for a hierarchical network, for example, in which ties flow along paths that do not loop back (e.g., an organizational chart) \citep{krackhardt1994}. The other relevant property of a metric space is the {\em triangle inequality}, which indicates that the distance between $i$ and $j$ can be at most the distance between $i$ and $k$ and $j$ and $k$, where $k$ is any other node in the network. This means that if, according to the latent space, $i$ is very likely to tie with $k$ and $j$ is very likely to tie with $k$, then $i$ and $j$ will also be likely to tie. The triangle inequality embeds transitivity into the structural assumptions underlying the LSM. Such an assumption would be inappropriate for a conflictual network, in which the enemy of your enemy is likely your friend \citep{cranmer2011inferential}. If the researcher's theory of network dependence is not precise enough to specify an ERGM and the structural assumptions associated with the LSM are not suitable in a given application, then QAP is the best option.

Model dependence is the term used to describe the conditions under which the conclusions drawn from a statistical modeling exercise vary across reasonable alternative specifications of the model \citep{ho2007}. For example, if a test of the hypothesis that nodes in a network exhibited homophily with respect to a given attribute exhibited different results when using LSM and QAP, the test would be model dependent. If the researcher's primary interest lies in testing hypotheses regarding the effects of exogenous covariates, and the researcher has identified plausible specifications of ERGM and/or LSM, is prudent to also evaluate hypotheses using QAP.  If results are consistent across ERGM, LSM, and QAP, that is a strong sign of robustness. If they vary, the researcher should endeavor to understand why and assure that the structural assumptions represented by ERGM and/or LSM are indeed appropriate.

Turning to dynamic models, the literature on political networks presents two modeling options -- SAOM and TERGM. \footnote{\cite{ward2013} propose a dynamic extension of the LSM. However, within-dyad autocorrelation is the only form of network dependence for which it accounts. Since it does not account for higher order forms of interdependence, we do not consider it to be a dynamic network model along the same lines as SAOM and TERGM, which both permit the modeling of arbitrary forms of inter-temporal dependence.} Both \cite{desmarais2012micro} and  \cite{leifeld2014comparing} present thorough discussions of the ways in which researchers can draw conceptual and empirical comparisons of SAOM and TERGM. First, in terms of empirical comparison, if the researcher does not have theoretical reason to prefer either SAOM or TERGM, it is possible to use either in or out-of-sample comparisons to evaluate which method provides a better fit. \cite{desmarais2012micro} and  \cite{leifeld2014comparing} provide extensive discussion and examples regarding the empirical comparison of SAOM and TERGM.

There are a few considerations that may allow the researcher to select between SAOM and TERGM based on theory. First, is it appropriate to assume that the network changes according to a process in which an actor considers changing one out-going tie at a time to stochastically optimize their objective function? If the answer to this question is yes, then the SAOM is a favorable option. Through making this assumption about the form of network change, it is possible to identify other features of the network dynamics, such as differentiating between homophily and influence. If the answer to this question is no, then TERGM is favorable because there are no such assumptions regarding the form of network change. Second, do the ties in the network and behavior co-evolve simultaneously? As we note above, the SAOM consists of two equations -- an objective function for tie formation and another for behavior (i.e., node attribute) evolution. The main advantage of using SAOM over estimating these functions separately is that, through the structural assumptions regarding change dynamics, it is possible to identify co-evolution dynamics that occur at a finer time interval than that at which the data was collected. Third, is there a reason to suspect non-stationarity in the time series of networks? Underlying the TERGM is an assumption that the series of networks under analysis has reached a stationary distribution -- a stable time-point-to-time-point process of network generation. In the SAOM it is assumed that change over time follows a stable/stationary process, but it is not assumed that the time series of networks has reached stationarity. Thus, if the network under study has recently come into formation -- suggesting that  more time may need to pass to reach stationarity, or there is some inexplicable shift in the network's structure over time, use of the SAOM would be favored.

\section{Frontiers in Network Modeling}

Scholars of government and politics are in a nearly unique position as far as access to data on ties between the actors they study. In every subfield of political science -- from political behavior research on campaign contribution networks of individuals to intra-institutional studies of interaction within legislatures or administrative agencies, to international studies of trade flows, alliances and migration --  data is regularly available and often open by public records mandate. Furthermore, this rich archival data often represents a census of the relevant population that covers several decades or even centuries. Since such varied and rich data is available, it is safe to assume that most established methods for modeling network data could be fruitfully applied to problems in political networks research. Furthermore, as the field of automated text analysis matures the volume and scope of network data available to political scientists is likely to explode. On the frontier of research we see great promise in the application of network methods to diverse dataset and in the generation of new political networks data. 

The frontier of statistical network analysis is also developing quickly. One area where the field is rapidly evolving is in techniques for analyzing weighted (e.g. valued-edge) networks: \citep{Wyatt:2010,Desmarais:2011plosone,Krivitsky:2011} have proposed methods for approaching this problem and software is now available to implement these techniques. Other recent developments that show promise and deserve further consideration in political networks research include the mixed membership stochastic blockmodel \citep{airoldi2009mixed}, exponential family random network models \citep{fellows2012exponential}, and relational event models for longitudinal data \citep{butts2008relational}, but these methods have, as of this writing, yet to permeate political science. 

In sum, political network analysis is developing rapidly in terms of the range of substantive problems to which network analysis is applied, the quality and volume of network data available to political scientists, and the statistical methods themselves. We think it likely that the landscape of political networks will look substantially different ten years after this writing than it does today.

\clearpage
\bibliographystyle{apsr}
\bibliography{dyad,psj,jss,LSM,qap,mmsbm}

\end{document}